\documentclass[10pt]{IEEEtran}
\usepackage{array}
\usepackage[switch,mathlines]{lineno}
\usepackage{soul}
\usepackage{graphicx}
\usepackage{cite}
\usepackage[cmex10]{amsmath}
\usepackage{algorithm}

\usepackage{algpseudocode}
\usepackage{fixltx2e}
\usepackage{CJK}
\usepackage[cmex10]{amsmath}
\usepackage{bm}
\usepackage{color}
\usepackage{amsmath,amsthm,amssymb,amsfonts}

\usepackage{arydshln} 

\usepackage{soul}
\usepackage{tikz}
\usepackage{booktabs}
\usepackage{tabularx}

\usepackage{cite}
\usepackage{hyperref}
\hypersetup{
     colorlinks   = true,
     linkcolor    = blue,
     citecolor    = red
}

\usepackage{xfrac}

\makeatletter
\newcommand*{\transpose}{%
  {\mathpalette\@transpose{}}%
}
\newcommand*{\@transpose}[2]{%
  \raisebox{\depth}{$\m@th#1\intercal$}%
}
\makeatother

\usepackage{scalerel}
\usepackage{tikz}
\usetikzlibrary{svg.path}

\usepackage{scalerel}
\usepackage{tikz}
\usetikzlibrary{svg.path}

\definecolor{orcidlogocol}{HTML}{A6CE39}
\tikzset{
    orcidlogo/.pic={
        \fill[orcidlogocol] svg{M256,128c0,70.7-57.3,128-128,128C57.3,256,0,198.7,0,128C0,57.3,57.3,0,128,0C198.7,0,256,57.3,256,128z};
        \fill[white] svg{M86.3,186.2H70.9V79.1h15.4v48.4V186.2z}
        svg{M108.9,79.1h41.6c39.6,0,57,28.3,57,53.6c0,27.5-21.5,53.6-56.8,53.6h-41.8V79.1z M124.3,172.4h24.5c34.9,0,42.9-26.5,42.9-39.7c0-21.5-13.7-39.7-43.7-39.7h-23.7V172.4z}
        svg{M88.7,56.8c0,5.5-4.5,10.1-10.1,10.1c-5.6,0-10.1-4.6-10.1-10.1c0-5.6,4.5-10.1,10.1-10.1C84.2,46.7,88.7,51.3,88.7,56.8z};
    }
}

\newcommand\orcidicon[1]{\href{https://orcid.org/#1}{\mbox{\scalerel*{
                \begin{tikzpicture}[yscale=-1,transform shape]
                \pic{orcidlogo};
                \end{tikzpicture}
            }{|}}}}

\usepackage{float}
\usepackage{multirow}
\usepackage{placeins}

\begin{document}


\title{ \LARGE{Community Resilience Optimization Subject to Power Flow Constraints in Cyber-Physical-Social Systems in Power Engineering}}
%
%

\author{Jaber Valinejad, Student member, IEEE, Lamine Mili, Life Fellow, IEEE
\vspace{-1.1cm}

\thanks{\hspace{-8pt}\underline{~~~~~~~~~~~~~~~~~~~~~~~~~~~~~~~~~~~~~~~~~~~~~~~~~~~}}
\thanks{The work presented in this paper was funded by the National Science Foundation (NSF) under Grant No. 1917308.

J. Valinejad and L. Mili are with the Bradley Department of Electrical and Computer Engineering, Virginia Tech, Northern Virginia Center, Greater Washington D.C., VA 22043, USA (email:{JaberValinejad,lmili}@vt.edu).}



}

\maketitle

\begin{abstract}

This paper develops a community resilience optimization method subject to power flow constraints in the Cyber-Physical-Social Systems in Power Engineering, which is solved using a multi-agent-based algorithm. The tool that makes the nexus between electricity generation on the physical side and the consumers and the critical loads on the social side is the power flow algorithm. Specifically, the levels of emotion, empathy, cooperation, and the physical health of the consumers, prosumers are modeled in the proposed community resilience optimization approach while accounting for the electric power system constraints and their impact on the critical loads, which include hospitals, shelters, and gas stations, to name a few. The optimization accounts for the fact that the level of satisfaction of the society, the living standards, and the social well-being are depended on the supply of energy, including electricity. Evidently, the lack of electric energy resulting from load shedding has an impact on both the mental and the psychical quality of life, which in turn affects the community resilience. 
The developed constrained community resilience optimization method is applied to two case studies, including a two-area 6-buses system and a modified IEEE RTS 24-bus system. Simulation results reveal that a decrease in the initial values of the emotion, the risk perception, and the social media platform effect factor entails an increase in load shedding, which in turn results in a decrease in community resilience. In contrast, an increase in the initial values of cooperation, empathy, physical health, the capacity of microgrids and distributed energy resources results in a decrease in the load shedding, which in turn induces an enhancement of the  community resilience.

\end{abstract}
\vspace{-0.2cm}
\begin{IEEEkeywords}
 Resilience; Community Resilience; Social Well-Being;  Cyber-Physical-Social System; Power Systems; Smart Grids; Social Computing; Power Flow;Load Shedding; Critical Loads.
\end{IEEEkeywords}
\vspace{-0.4cm}

\section*{Nomenclature}

\begin{tabular}{p{0.7cm} p{7.2cm}}
\multicolumn{2}{c}{\textit{1- Indexes}} \\
t  & Index for time \\
n/m  & Index for bus (N is the total number of buses)\\
\multicolumn{2}{c}{\textit{2- Social science Variables}} \\
$M^{e}_{tn}$  & The level of emotion (fear) in each bus\\
$M^{r}_{tn}$  & The level of risk perception in each bus\\
$M^{c}_{tn}$ & The level of cooperation in each bus\\
$M^{a}_{n}$ & The level of empathy in each bus\\
$M^{p}_{tn}$ & The level of physical health in each bus \\
$S_{t}$  & The level of social well-being of a community \\
\multicolumn{2}{c}{\textit{3- Power flow variables}} \\
$\alpha_{nt}$/
$\beta_{nt}$        &  The Load shedding  of consumers, prosumers/critical loads\\
$P_{nmt}$           &  The electricity transferred between two buses \\
$\theta_{nt}$       &  The voltage angle   \\
$P^{der}_{nt}$      &  The electricity produced by Distributed energy resources  \\  
$P^{u}_{nt}$  &  The electricity produced by utilities   \\

\end{tabular}
\begin{tabular}{p{0.7cm} p{7.2cm}}

$P^{mg}_{nt}$       &  The electricity produced Micro Grids    \\

$P^{cl}_{nt}$       &  The electricity consumed by critical Loads   \\
$P^{d}_{nt}$        &  The electricity consumed by consumers and prosumers \\

$\overline P^{l}_{nm}$    & The capacity of transmission line  \\

$\overline P^{der}_{n}$   & The capacity of distributed energy resources    \\
$\overline P^{mg}_{n}$  & The capacity of micro grids \\
$\overline P^{u}_{n}$  & The capacity of generation unit \\
$\textit{3- Cyber variables}$\\
$N^{m}_{t}$ & The level of the related and negative news of mass media\\
\end{tabular}

\vspace{-0.6cm}

\section{Introduction}

When a social community is exposed to natural and human-induced disasters, it faces a variety of emotional and physical stresses and strains, which may result in physical and financial losses and loss of life. The question is hence the following: What should that community do to better face a given disaster and decrease the losses that it may experience? To address this question, the resilience of a community must first be defined and characterized by relevant metrics, whose levels must be assessed and enhanced. The features of a social system include emotion, empathy, risk perception,  cooperation, social well-being, and community resilience. In this paper, community resilience is defined as the ability of a community to bounce back and recover from a given class of severe disturbances \cite{mili2018,Cutter2008}. One social system feature that has an important impact on community resilience is social well-being, whose modeling and assessment require an interdisciplinary approach, integrating knowledge and ideas from a variety of disciplines such as neuroscience, social and cognitive psychology, artificial intelligence, cognition, multimedia development, engineering, and healthcare \cite{calvo2014affect}. Social well-being consists of mental well-being and physical well-being. In this paper, we measure the level of mental well-being  by the level of fear, which is of course affected by psychological and mental characteristics such as cooperation, empathy, and risk perception. In contrast, we measure the level of  physical well-being by the level of physical health. Using these metrics, we investigate how the availability of electricity impacts the community resilience. \

\begin{figure}[h]
\centering
\includegraphics[width= 0.8 \columnwidth]{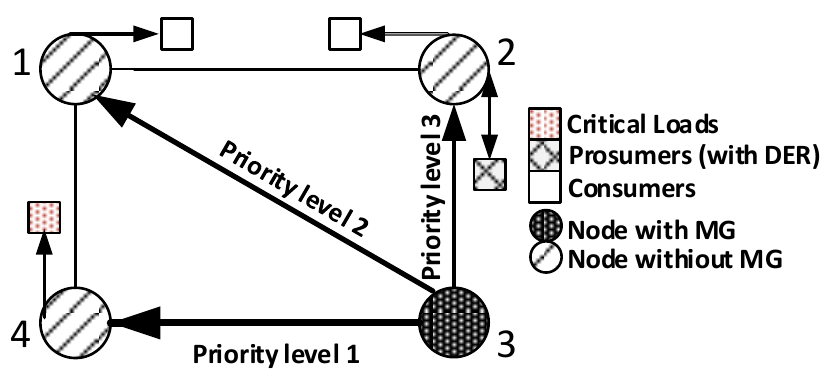}
\vspace{0.25cm}
\setlength{\abovecaptionskip}{-15pt}
\caption{ Community resilience maximization subject to power flow constraints in the CPSS-PE. Node 1 includes consumers. Node 2 includes both consumers and prosumers. Node 3 includes a Microgrid (MG). Node 4 includes critical loads. 
}
\vspace{-0.3cm}
\label{fig:Fig_355}
\end{figure}

While the availability of electricity, as the main type of energy sources, directly affects the physical quality of life, the life expectancy, the human development and health, just to name a few \cite{carvallo2017sustainable}, the risks associated with its shortage are not always promptly visible. Evidently, its shortage or unavailability threatens human lives and makes people mentally unsatisfied with the power suppliers, e.g., utilities, retailers, and the government.  Hence, it is essential to consider the community’s social well-being in Cyber-Physical-Social Systems in Power Engineering (CPSS-PE), before, during, and after the striking of a disaster. Evidently, in case of shortage of electricity, the critical loads must be supplied with the highest priority. Furthermore, experience has shown that the level of the social well-being is higher if there is some supply of electricity as compared to the case where there is no supply of electricity, especially during a disaster \cite{rubin2019behavioural}. Figure~\ref{fig:Fig_355} displays in a graphical manner a simple example of a four-bus system, where only consumers are connected to Bus 1, consumers and prosumers are connected to Bus 2, a microgrid is connected to Bus 3, and critical loads are connected to Bus 4. When an emergency occurs, the microgrid of Bus 3 supplies first the critical loads of Bus 4 with a priority level 1 by switching on its circuit breaker while the circuit breakers of the other loads are turned off. If the microgrid has enough electric energy, it supplies then the consumers of Bus 1 with a priority level 2. Finally, it supplied the consumers and prosumers of Bus 2 with a priority level 3. \par
\vspace{-0.8cm}

The tool that makes the nexus between the electricity generation on the physical side and the consumer and the critical loads on the social side is the power flow algorithm. The latter is the essential tool for the long-term and operational planning of a power system \cite{ma2017pricing,zhang2019event}. Owing to its importance, the power flow method in cyber-physical systems has already been studied for various applications, but without considering the social science aspects \cite{abdi2017review,tostado2019robust}. Hence, we are motivated to propose the socio-technical power flow  \footnote{Socio-technical system is a joint system referring to the interaction between human behavior and community's complex infrastructure such as power systems.\cite{long2018socioanalytic}.} in the CPSS-PE. The socio-technical power flow algorithm is the main tool for the analysis of a power system in the CPSS-PE. In this algorithm, the loads that impact the most the community resilience and that provides the highest community satisfaction need to be given the highest priority of supply. These loads must be supplied according to the capacities of the microgrids, the distributed energy resources (DERs), and the transmission lines \cite{ye2016game}. In reality, we face a more sophisticated power system than the example presented in Figure~\ref{fig:Fig_355}. Consequently, the socio-technical power flow becomes a challenging problem to solve due to the numerous technical and social constraints.\par

\begin{figure}[h]
\centering
\includegraphics[width= 1 \columnwidth]{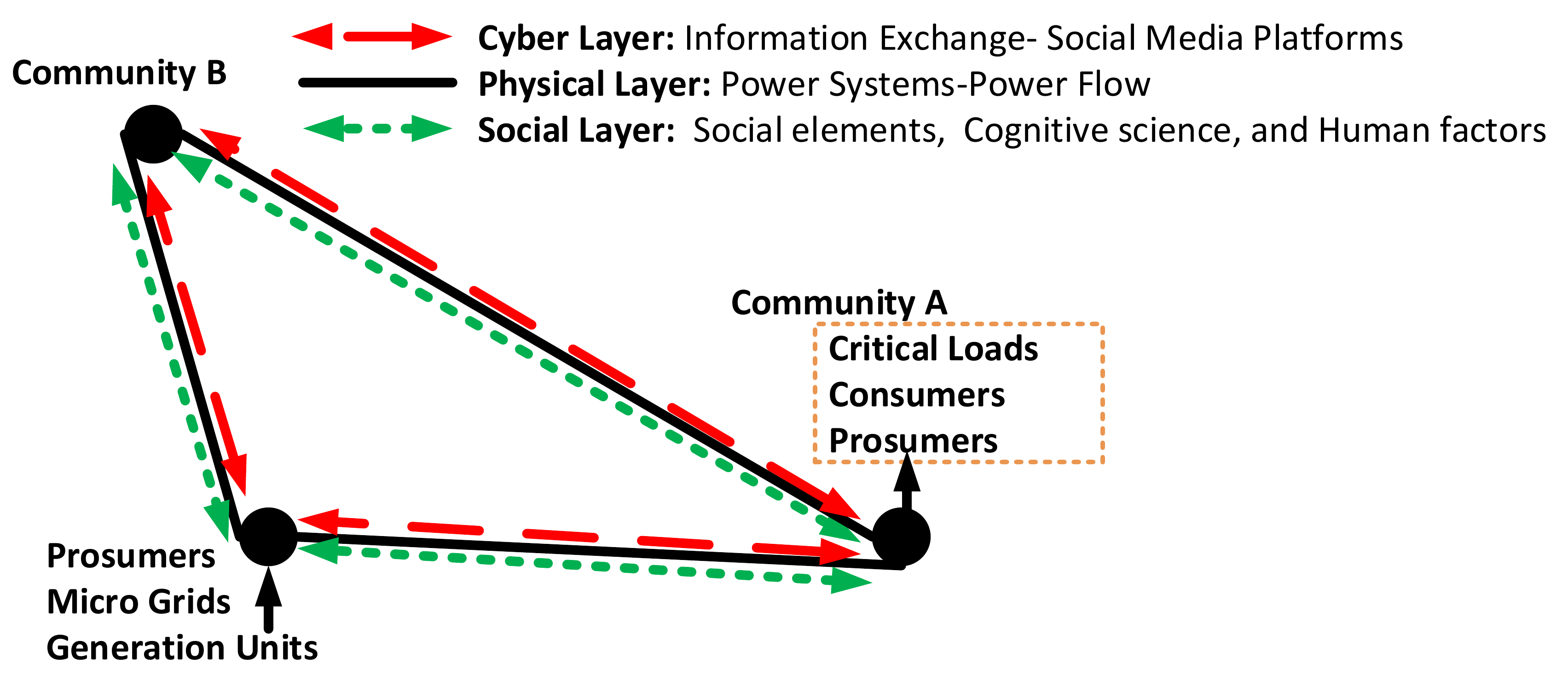}
\setlength{\abovecaptionskip}{-15pt}
\caption{Diagram displaying a cyber layer, a physical payer, and a social layer of our CPSS-PE model. The cyber layer comprises the social media platform that performs the exchange of information. The physical layer comprises the power system.  The social layer  comprises social elements and human factors. }
\vspace{-0.2cm}
\label{fig:Fig_34}
\end{figure}

This paper is an extension of our previous work described in  \cite{jaber2019b,jaber2019}, which does not account for the power flow constraints, the load shedding, and the electricity shared during disaster that is limited by the capacity of the transmission lines. Specifically, we will address here the following question: How to maximize community resilience subject to power flow constraints in CPSS-PE? To this end, our CPSS-PE model considers the social perspectives of engineering systems, where the human and social dynamics are considered as an integral part of any effective cyber-physical system design and operation \cite{kang2018managing}. It accounts for the tight conjoining and coordination between the physical world, the cyber world, and the social world, as first proposed by Karl Popper \cite{zhang2018cyber}. Figure~\ref{fig:Fig_34} displays a diagram of the CPSS-PE that we developed, which consists of a cyber layer, a physical payer, and a social layer. The cyber layer comprises the social media platform that performs the exchange of information. The physical layer comprises the power system.  As for  the social layer, it comprises social elements and human factors described by the social and cognitive science and psychology. In summary, our CPSS-PE models the cyber-physical-social dependence among prosumers, consumers,  microgrids, power utilities, and mass media platforms.
\vspace{-0.4cm}

This paper addresses the following sub-questions:
\begin{itemize}
\item How can we model the social well-being of a society subject to power flow constraints?
\item How can we demonstrate the effect of the level of cooperation of DERs and prosumers on the sharing of electricity? 
\item How can we model the impact of load shedding and mass media platforms on the level of fear, cooperation, risk perception of consumers and prosumers, and social well-being?
\item What is the effect of load shedding on the emotion, risk perception, physical health, empathy, and cooperation, and hence on community resilience in CPSS-PE?
\item What are the effects of the capacity of microgrids and DERs on the  social characteristics?
\item What are the effects of the mass media platforms on the load shedding?
\item What are the effects of the availability of electricity on the physical and mental well-being?
\item How do the emotions, risk perception, physical health, and cooperation dynamically change during a day?
\item How does the amount of load shedding that consumers, prosumers, and critical loads may experience change during a day?
\item How to maximize the community resilience under a limited amount of electric energy by minimizing the load shedding that consumers, prosumers, and critical loads may experience while satisfying the power flow constraints? 
    \end{itemize}

We answer the three first sub-questions in Section III. The other sub-questions are answered in Section IV. The community resilience optimization method subject to power flow constraints is implemented in two different case studies. This model is verified by the soft validation and sensitivity analyses. The aim of the first case study, i.e., two-area 6-buses system, is to investigate the effect of the level of empathy, the amount of mass media effect factor, the DERs capacity, the microgrid capacity, the initial value of fear, the social cooperation, the risk perception, and the physical health on consumer load shedding, prosumers, and critical loads, the reporting of negative news by the mass media platform, the mental well being, the physical well being, the social well-being, and the community resilience. To reach our aims, we provide the results for 24 distinct scenarios. The sub-questions 4-7 are elucidated in this case study. The second case study is carried out on the modified IEEE RTS 24-bus system. It intends to provide a dynamic change of load shedding of consumers, prosumers, critical loads, levels of fear, cooperation, risk perception, and community resilience for 24 hours. The three last sub-questions are clarified in this case study. \par
The remainder of this paper is structured as follows. Section II introduces the social computing and social characteristics considered in our model. It also discusses the social behavior and emotion, the Barsade theory, the Fredrickson theory, the amplification model, and the absorption model. In addition, it provides the definition and stresses the importance of cooperation, empathy, risk perception, social well-being, critical loads, power flow, and load shedding. Section III deals with the community resilience optimization problem subject to power flow constraints in CPSS-PE. It also explains the inputs and the outputs of the proposed model as well as cyber-physical-social dependence in the proposed multi-agent-based model. Section IV discusses the results of the proposed method for the first case study carried out on the two-area 6-bus system. Section V discusses the results of the proposed method for the second case study carried out on the modified IEEE RTS 24-bus system. The conclusions are provided in Section VI.
\vspace{-0.4cm}

\section{Social Computing and Collective Behavior}
In our community resilience optimization subject to power flow constraints, the novelty of the approach resides in the modeling of the human behavior in CPSS-PE. However, there are a number of papers dealing with the modeling of the social behavior in cyber-social systems. One approach to model social behavior is multi-agent-based modeling of teamwork cooperation \cite{rishwaraj2016heuristics}. It is used in various complex system modeling and real-world applications, such as  transportation systems, social-economic systems, energy systems, and  online friendship network systems (e.g., Facebook and Twitter) \cite{wang2013community}. Tan $et$ $al.$ \cite{tan2016evolutionary} model the dynamic of collective behavior by using game theory while  considering the effect of the social norms and cultural trends. They also discuss in details a number of collective behavioral patterns, influenced by a variety of conflicts in social networks, such as behavioral flocking, collapse, and oscillation. Ning $et$  $al.$ \cite{ning2017collective} model the collective behavior by using the nearest neighbor rule. Meo $et$ $al.$ \cite{de2014trust} show that human factors, such as emotion, risk perception, and cooperation, have a profound effect on the dynamics of the collective behavior in social networks. Giraldo and Passino \cite{giraldo2015dynamic} discuss the dependence between the cohesiveness of the group and its performance. They show that a large number of connections between individuals reduces the cohesiveness of the group and its performance. In other words, a decentralized communication network has better performance than a centralized one. Yu $et$ $al.$ \cite{yu2014collective} consider the social norms and conventions to predict collective behavior by applying a  multi-agent-based model.\par
In our model, we consider emotion, risk perception, empathy, cooperation, and social well-being and their effects on the community resilience. Their definitions and the meaning of their numerical values are provided in Table~\ref{tab:FCC}. There is a widely-held belief that emotion is the core characteristic of group behavior; consequently, modeling group emotion is of high importance. So, we first discuss emotion and then we discuss the aforementioned social features.\par
\vspace{-0.1cm}
\begin{table*}
\centering
\scriptsize
\setlength{\abovecaptionskip}{-2pt}
\caption{Definition of the social system traits, and the meaning of their numerical values. The social system features include emotion, risk perception, empathy, cooperation, social well-being, and community resilience. The metrics of these features are assumed to take values in the interval [0 1].}

\begin{tabular}{|p{2.3cm}|p{6cm}|p{7cm}|p{1.2cm}|}
\hline
Characteristic&  Definition & Value (between [0,1]) & References\\
\hline
Emotion &  The Fear felt by an individual during a disaster  &  0 means the agent does not have any fear, 1 means the highest level of fear & \cite{Barsade1998,silver2002nationwide,6472097,bonanno2007predicts,bonanno2001loss,xu2019crowd}\\

\hline
Risk Perception & The feeling that an agent perceives that he/she is in jeopardy &  0 means the agent does not feel any risk, 1 means the highest level of perceived risk & \cite{ping2018modeling ,shin2018human,allen2019culture}\\

\hline
 Empathy & The experience of other people' emotion and thoughts  &  0 means there is no compassionate empathy between two agents, 1 means the highest level of empathy exist & \cite{banfield2012role,hojat2016empathy,peters2014compassion,rochat2002various}\\
\hline
Cooperation  & Willingness to work unitedly on a particular number of task and sharing resources, information, and experience that aimed to common goal and objective &  0 means the agent does not have any willingness to cooperate, 1 means the highest level of cooperation the agent has & \cite{axelrod1981evolution,7583665,parkash2015cooperation,fehr1999theory,nowak2006five}\\
\hline
Social Well-being  & The experience of both physical and mental health. It directly influence the quality of life  & 0 means the society does not have any well-being, 1 means the highest level of social well-being the community has & \cite{dodge2012challenge,burke2010social,calvo2014affect}\\
\hline
Community Resilience  & The ability of a community to bounce back and recover from a given class of extreme disturbances & 0 means the community is not resilient    to  a  given specific class of  disturbances, 1 means the highest level of community resilience  the society has &  \cite{Cutter2008,mili2018} \\
\hline
\end{tabular}
\label{tab:FCC}
\vspace{-0.6cm}
\end{table*}
\vspace{-0.5cm}

\subsection{Emotion}
In addition to logical intelligence, emotional intelligence is part of human intelligence \cite{zheng2018emotionmeter}. Emotions are complex psycho-physiological processes that are controlled by many internal and external factors \cite{zheng2018emotionmeter}. Human emotion plays a crucial role in both human-human and human-machine interaction \cite{zhang2018spatial}. Emotion in social intelligence is also important. Ficocelli $et$ $al.$ \cite{ficocelli2015promoting} provide a model for the human-robot interaction by using robotic emotional behavior. In addition, there are various approaches to reorganize and classify emotional behaviors. Emotion recognition by Electroencephalogram (EEG) is proposed in \cite{zhang2018spatial, zheng2018emotionmeter,li2019multisource}. Furthermore, Deb $et$ $al.$
discuss emotion classification. In our model, to apply emotion, we make use of the Barsade theory, the broaden-and-build theory, the amplification model, and the absorption model. Finally, we explain these concepts in the summary.

\subsubsection{Group Emotion: Barsade Theory}

Barsade $et$ $al.$ \cite{Barsade1998} propose a top-down and a bottom-up approach to model group emotion. On one hand, in the top-down approach, emotion flows from the group level to the individual level so that the emotion raises at the group level is felt by each person (or agent). On the other hand, in the bottom-up approach, individual emotion can influence the group emotion. It is evident that in the latter approach, the group emotion is formed by the combination of the feeling of each member (or agent).

\subsubsection{Upward Emotional Well-Being: Fredrickson Theory}

One important question, which is pivotal for the social network emotion, is the following: How do positive and negative emotions influence the agents? Fredrickson $et$ $al.$ \cite{Fredrickson2002} answer this question by proposing a broaden-and-build theory (or Fredrickson theory). Based on this theory, negative affect (emotion) restricts the individual’s thoughts and actions; positive emotion, on the contrary, broadens the set of thoughts and actions of people. According to this theory, joy induces a feeling to play, contributing to physical, socio-emotional, and intellectual resources (skills) so that they lead to brain development. Correspondingly, interest leads to motivation to explore, causing physical, social, intellectual, and psychological skills. As a result, an increase in personal or agent’s resources is the consequence of positive emotions. According to the  broaden-and-build theory, two new conceptions, i.e., upward spirals and downward spirals, are introduced. In upward spirals theory, it is a belief that positive emotions broaden thought-action proceedings, attention, and cognition, both at present and in the future. Also, based on positive statuses such as well-being, optimism, and success, prognosticate global biases in accordance with widened attention. On the other hand, in downward spirals theory, negative status, such as anxiety, depression, and failure, anticipate local prejudices according to narrowed focus.

\subsubsection{Absorption Model - a Multi-Agent-Based Model for Group Emotion}

To model the emotion of social networks, computational models are used. According to the social neuroscience, emotion can be considered as a collective feature of the group so that the emotion of an agent  can form the feelings, thoughts, and behavior of other agents.


In the absorption model, the bottom-up conception based on Barsade theory is used  \cite{Bosse20092}. According to this approach, the team emotion is equal to the sum of its parts in which the group emotion is influenced by homogeneity, heterogeneity, and the mean emotion of agents within the group. This model is appropriate in some situations where the simulation of the emotion dynamics of the agents is important.

\subsubsection{Amplification Model}
The amplification model to model the emotion of social networks is based on Fredrickson theory, i.e., the broaden-and-build theory, including upward and downward emotional spirals. If there is no outside event or disaster, the absorption model can be appropriate. On the other hand, the amplification model is for cases where there are sudden events and obstacles in the group, emergency, and the factors outside the group that can influence the group emotion. Here, the community resilience planner may use both approaches.
\vspace{-0.35cm}

\subsection{Cooperation}
We use a multi-agent-based model to examine the social behavior. In a multi-agent system, the success or failure in accomplishing an objective is highly dependent on the cooperation between the agents \cite{rishwaraj2016heuristics}. According to the World Peace Through Technology Organization (WPTTO), cooperation between agents induces much more benefits than competition \cite{ER19}. Hence, modeling cooperation and its effect on social behavior are of high importance. Guan $et$ $al.$ \cite{guan2019cooperation} propose a cooperation model from the multiple social networks. Shao $et$ $al.$ \cite{shao2020bipartite} discuss the simultaneous impact of cooperation and competition. Besides, different factors influence the level of cooperation among the social group. The main feature for teamwork cooperation is trust between agents \cite{rishwaraj2016heuristics}. De $et$ $al.$ \cite{de2014trust} emphasize the importance of mutual trustworthiness between agents to cooperate and to form a social group. In addition, for efficient team cooperation, there is a need for a social connection among agents \cite{wang2016toward}. Wang $et$ $al.$ \cite{wang2016toward} discuss the effect of the selfish agent in social networks on collective behavior.

\vspace{-0.5cm}

\subsection{Empathy}
The key element in establishing meaningful and effective social relationships is empathy. Emotional support needs to be an empathetic communication where one understands the emotional state of other people \cite{bickmore2010empathic,salminen2019evoking}. Empathy is assumed to be an emotional and spiritual feature that makes individuals understand other people. Empathy can be taught and shared, and it increases social cooperation. Unfortunately, empathy in the United States has declined by 50\% during the past 40 years, and the steepest decline happened during the last ten years \cite{ER19,ERB}. This decline in  empathy reduces community resilience . To increase empathy among people, benevolent technologies, and Code4Peace program as smartest approaches to social change are recently proposed. Benevolent technologies include peace software\footnote{Peace software are tools and platforms that aim to make peace in the community and to increase the awareness of global interdependency \cite{ER19}.}, media technology, communications technology, compassion, stories, peace games, bicycle power, and green technology. In addition, Code4Peace is a program that encourages programmers and peace workers to collaborate. Code4Peace aims to create peace by making practical and valuable software.
\vspace{-0.5cm}

\subsection{Risk Perception}
One of the natural behavior of people when they face disaster is the feeling that they are in danger due to their dynamic interaction with the environment. This helps them to take actions aimed at dealing with the situation and the incident. With an increase in the uncertainty of a disaster, people tends to perceive a higher risk than it is in reality. This is because they may be at greater risk otherwise. People without previous experience, they cannot unusually evaluate the risk of a hazard as reliably as someone with prior experience. Consequently, they are exposed to greater danger.
Risk perception can be subjective-based or objective-based. 
Ping $et$ $al.$ \cite{ping2018modeling} model subjective risk perception of a driver by using deep learning while Shin $et$ $al.$ \cite{shin2018human} propose a human-centered approach to model risk perception. Different people perceive different risks when they face  different types of disasters. Factors such as judgment, situational awareness, experience, culture, and cognition influence how people evaluate the danger of a situation \cite{ping2018modeling ,allen2019culture}. The risk perceived by individuals during disaster form the public risk perception and the social interaction and communication. Allen $et$ $al.$ \cite{allen2019culture} propose the psychological model for public risk perception under extreme heat events, the major weather-related cause of death in the United States, and flooding.
\vspace{-0.35cm}

\subsection{Social Well-Being and Community Resilience}
Social well-being requires an interdisciplinary approach, integrating knowledge and ideas from disciplines such as neuroscience, social and cognitive psychology, artificial intelligence, cognition, multimedia development, engineering, and healthcare \cite{calvo2014affect}.
The social well-being includes the mental and physical well-being. In our model, we consider the inverse of the level of emotion (fear) of a society as mental well-being. This means that the less concern, the more mental well-being. On the other hand, we hold the physical health of the community as a physical well-being. When a society faces a disaster or an extreme event, its social well-being is affected, especially when there are losses. The social well-being is therefore considered as the main feature that affects the community resilience. In this paper, the objective is to maximize the social well-being during a disaster while reducing the negative social impact of a shortage of electricity via a better cooperation among prosumers, consumers, and critical loads for sharing scarce resources such as electric energy produced by microgrids and DERs subject to power flow constraints, which in turn enhance community resilience. The reader is referred to Section III for further details.
\vspace{-0.35cm}

\subsection{Power System in Social Science}
\vspace{-0.05cm}
In addition to environmental and economic issues, the use of energy indicators are relevant to social issues \cite{kemmler2007energy,wijayasekara2014fn}. Consumers of electricity and critical loads are part of the social systems. Arto $et$ $al.$ \cite{arto2016energy} clarify the dependence between human development index, welfare, and electricity. By providing electricity to humans based on their needs and their satisfaction, living standards are improved \cite{goldemberg1985basic}. Hence, a reliable supply of electricity to a community is essential. By contrast, shortage of electricity and load shedding degrade both the mental and physical quality of life. Physiological changes as a function of electrical energy consumption are not immediately  manifested \cite{alam1998revisited}. Alam $et$ $al.$ \cite{alam1991model} present a model for the physical quality of life as a function of per capita electrical energy consumption. The tool that makes the connection between electricity generation on the physical side and consumer and critical loads on the social side is power flow calculation. Hence, we first discuss the latter. Then, we discuss critical loads and load shedding.

\subsubsection{Critical Loads}
Critical loads must be supplied with the highest priority, an action that significantly impacts the level of community resilience. They consists of hospitals, operating theaters in hospitals, data centers, information and communication technology centers, ultraviolet lights in water treatment plants, radar equipment for airports, booster systems in pipeline applications, and emergency lighting systems.
 
\subsubsection{Power Flow Equations}
Derived from Ohm’s law and Kirchhoff’s current and voltage law, power flow equations are used for deriving all the functions of an energy management system \cite{li2019distributed}. These functions include static state estimation, optimal power flow, contingency analysis, power system planning, unit commitment, and reliability assessment \cite{li2017approximate}. In the power flow model, active and reactive power injections at each bus are expressed as nonlinear equations of the bus voltage magnitudes and voltage phase angles \cite{yang2018general}. Various power flow models have been proposed in the literature   \cite{li2017approximate,xu2019probabilistic,yang2018general,shchetinin2018construction,tostado2019robust}. These models may be based either on logarithmic transform, or on adaptive polynomial chaos-ANOVA method, or on a general representation of independent variables, or on constructing inner and outer linear approximations, or on Bulirsch–Stoer method. Power flow methods for power distribution systems are reviewed in Yang $et$ $al.$ \cite{yang2018general}. 

\subsubsection{Load Shedding}
Rolling blackout in electric power grid, also known as rotational load shedding, is an emergency control tool initiated by electric utilities aimed at curtailing the  excess of load with respect to the power generation due to unplanned failures or an unexpected large increase of the load for blackout prevention \cite{xu2016optimization}. In other words, rolling blackouts are the last resort measure employed by electric utilities to prevent overloading, instability, and system collapse of the power grid \cite{nourollah2018coordinated}. The California electricity crisis of 2000-2001 \cite{CAL2002}, and Western Victoria and South Australia incidents on 24 and 25 January 2019, respectively, \cite{AEMO2019}, are real examples of rolling blackouts that are due to unplanned system inefficiencies, the lack of maintenance of generating units and power transmission and distribution systems, increased population, and improved living standards \cite{xu2016optimization}.
\vspace{-0.45cm}




\section{Community Resilience Optimization Subject to Power Flow Constraints}

There are three different types of inputs to our multi-agent-based model, namely cyber-based, physical-based, and social-based inputs. The cyber-based inputs consist of the social media effect factor ($\zeta^{m}$). Physical inputs include the capacity of the transmission line ($\overline P^{l}_{nm}$), the capacity of distributed energy resources  ($\overline P^{der}_{n}$), the capacity of microgrids ($\overline P^{mg}_{n}$), and the generation unit capacity ($\overline P^{u}_{n}$). There are two different types of social-based inputs, including diffusion-based and social-initial-based inputs. Regarding the inputs of the diffusion features, they consists of the emotion contagion as a diffusion factor ($\gamma^{e}_{ij}$), the assumed initial value of fear ($M^{e}_{(t=1)i}$), the risk perception ($ M^{r}_{(t=1)i}$), the cooperation ($M^{c}_{(t=1)i}$), the empathy ($M^{a}_{i}$), the physical health ($M^{p}_{(t=1)i}$), and the social well-being ($S_{(t=1)}$).\par
As for the outputs of our multi-agent-based model, they consist of  cyber-based, physical-based and social-based outputs. Cyber-based output include the related and negative news propagated in the mass media platforms because of load shedding ($N^{m}_{t}$). Physical-based Variables and outputs include the Load shedding of consumers/critical loads ($\alpha_{nt}/\beta_{nt}$), the electricity transferred between two buses ($P_{nmt}$), the voltage angle ($\theta_{nt}$), the electricity produced by DERs ($P^{der}_{nt}$), the electricity produced by utilities ($P^{u}_{nt}$), the electricity produced by microgrids ($P^{mg}_{nt}$), the electricity consumed by critical Loads ($P^{cl}_{nt}$), and the electricity consumed by the consumers and the prosumers ($P^{d}_{nt}$). As for the social-based outputs, they comprise the incremental changes of fear ($M^{e}_{(t\neq1)i}$), the risk perception ($ M^{r}_{(t\neq1)i}$), the cooperation ($M^{c}_{(t\neq1)i}$), the physical health ($M^{p}_{(t=1)i}$), the social well-being ($S_{(t\neq1)}$), and the social mental and physical well-being.\par
\vspace{-0.5cm}


\subsection{Description of the Constrained Optimization Model}

The social well-being of a CPSS-PE is formed of the social mental well-being and the social physical well-being of a set of  consumers, prosumers, and critical loads. In this section, we plan to maximize the social well-being, $S_{t}$, that is, the community resilience, subject to a set of cyber-physical-social constraints. Formally, we have

\vspace{-0.25cm}
\footnotesize
\begin{equation}
\textbf{Max} \sum_{t}S_{t}
\label{eq:15} 
\end{equation}
\normalsize
\vspace{-0.3cm}
subject to
\noindent
\footnotesize
\begin{eqnarray}
S_{t}=\frac{1}{N} (  \sum_{n}\eta(\zeta^{e}(1-M^{e}_{tn})+\zeta^{ecl}(1-M^{ecl}_{tn}) ) \nonumber\\
+\sum_{n}(1-\eta)(\zeta^{p} M^{p}_{tn}+ \zeta^{pcl}P_{tn}^{cl} )).
\label{eq:15} 
\end{eqnarray}
\normalsize

and eleven other equality or inequality constraints that are defined next.

In (2), the first term of the summation is the social mental well-being while the second term is the social physical well-being. The well-being coefficients are contained in the set $L_{WC}$=\{$\eta,\zeta^{e},\zeta^{ecl},\zeta^{p},\zeta^{pcl}$\}. The reader is referred to the nomenclature for the definitions of the variables and their indices shown in (2). The critical loads respectively influence the mental and the physical well-being via 

\footnotesize
\begin{equation}
M^{ecl}_{tn}=\varpi^{e}(1-\beta_{tn})
\label{eq:15} 
\end{equation}

\noindent
\begin{equation}
P^{cl}_{tn}=\varpi^{p} \beta_{tn}
\label{eq:15} 
\end{equation}
\normalsize

where $\varpi^{e}$ and $\varpi^{p}$ are respectively the mental and the physical coefficients. The load shedding variable, $\beta_{tn}$, is constrained to takes values between 0 and 1,  
0 $\leq$ $\beta_{tn}$  $\leq$ 1.

The optimization given by (1) is subject to a second set of equality constraints, which are given by the human psychological dynamics. These are the dynamical changes of the level of emotion (fear) of consumers and prosumers in CPSS-PE. they are expressed as

\footnotesize
\begin{equation}
M^{e}_{(t+1)n}=\gamma^{e}_{tn}(f(\hat{M}^{e}_{tn},M^{e}_{tn})-M^{e}_{tn}) \varkappa^{t} +M^{e}_{tn} , 
 \label{eq:1} 
\end{equation}
\normalsize
where $\varkappa^{t}$ denotes the time coefficient such that $\varkappa^{t} \leq \frac{1}{n-1}$ as indicated in \cite{Bosse20092} and where 

\footnotesize
\begin{equation}
\gamma^{e}_{tn}=
\frac{\sum_{m} \gamma^{e}_{nm} M^{e}_{tm}}{\sum_{m} \gamma^{e}_{nm}},
 \label{eq:2} 
\end{equation}
\normalsize
\vspace{-0.7cm}

\footnotesize
\begin{eqnarray}
f(\hat{M}^{e}_{tn},M^{e}_{tn})=\eta^{e}[M^{r}_{tn}(1-(1-M^{e}_{tn})(1-\hat{M}^{e}_{tn}))
\nonumber \\+(1-M^{r}_{tn})(\hat{M}^{e}_{tn}M^{e}_{tn})]+(1-\eta^{e})\hat{M}^{e}_{tn},
\label{eq:3} 
\end{eqnarray}
\normalsize
\vspace{-0.7cm}
\footnotesize
\begin{eqnarray}
\hat{M}^{e}_{tn}=w^{ee} (\frac{\sum_{m} \gamma^{e}_{tnm} M^{e}_{tm}}{\sum_{m} \gamma^{e}_{tnm}})
+W^{ce}(1-M^{c}_{tn}) \nonumber \\
+W^{pe}(1-M^{p}_{tn})+W^{\alpha e}(1-\alpha_{tn})+W^{me}N^{m}_{t},
\label{eq:4} 
\end{eqnarray}
\normalsize
where $\alpha_{tn}$ denotes the load shedding variable, which is constrained to takes values between 0 and 1, 0 $\leq$ $\alpha_{tn}$  $\leq$ 1. 
Here, $\gamma^{e}_{tn}$ denotes the weighted emotion contagion of each agent based on the bottom-up approach,  which is also considered as the speed of the dynamic change of the total emotion strength of a consumer or a prosumer of a group receiving the emotion of the other consumers and prosumers within that group. As for $f(\hat{M}^{e}_{tn},M^{e}_{tn})$, it denotes the  amount of the impression of the inter- and the intra-agent factors through the absorption and the amplification model. Akin to the absorption model based on the Barsade theory,  $\hat{M}^{e}_{tn}$ denotes the amount of emotion of an agent influenced by the emotion of the other consumers and prosumers, which account for the inter-agent impacts  \cite{natalie_13}. Here, the term,  $[M^{r}_{tn}(1-(1-M^{e}_{tn})(1-\hat{m}^{e}_{tn}))+(1-M^{r}_{tn})(\hat{M}^{e}_{tn} M^{e}_{tn})]$, is associated with the amplification model based on the Fredrickson theory. This model consists of two different terms that are related to an upward and a downward emotional spiral, respectively. In (8), the weighting factors are contained in the set $L_{W}$=\{$w^{ee},W^{ce},W^{pe},W^{\alpha e}$\}. 

Note that $\hat{M}^{e}_{tn}$ is influenced by the social-social dependence including the emotion of the other agents ($w^{ee} (\frac{\sum_{m} \gamma^{e}_{tnm} M^{e}_{tm}}{\sum_{m} \gamma^{e}_{tnm}})$), its cooperation ($W^{ce}(1-M^{C}_{tn})$) \cite{rand2014positive,lebow2005reason,levine2018signaling,kjell2013exploring}, and agent's physical health ($W^{pe}(1-M^{p}_{tn})$) \cite{ohrnberger2017relationship}. In addition to the social-social dependence, the level of panic is contingent on the physical-social dependence,i.e., the load shedding of consumers and prosumers ($W^{\alpha e}(1-\alpha_{tn})$) and the cyber-social dependence, i.e., the mass media ($W^{me}N^{m}_{t}$) \cite{paperasli}. It is prevalent for users to follow news or events conveyed by the  social media platforms, such as Twitter, Facebook, Sina Weibo, WeChat, and energy media \cite{zhao2015granular}. They use these social media services to share their emotions and thoughts \cite{pang2019fast}. The dynamic change of the level of the related and negative news of mass media is given by

\footnotesize
\begin{equation}
N^{m}_{t}= \zeta^{m}[\zeta^{'e}(1-\alpha_{tn})+\zeta^{'ecl}(1-\beta_{tn})]
\label{eq:5}
\end{equation}
\normalsize
Here, the mass media news are directly related to the load shedding of consumers, prosumers, and critical loads. $\zeta_{m}$ is the effect Coefficient. Note that in (9), we have disregarded the effect of the fake, exaggerated, or tendentious news. If the level of satisfaction of a consumer at a bus is desired to be high, we can set the level of emotion in (5) accordingly. 
The optimization given by (1) is subject to a third equality constraint, which is the dynamic change of the level of risk perception of consumers and prosumers in CPSS-PE given by
\vspace{-0.6cm}

\footnotesize
\begin{eqnarray}
 M^{r}_{(t+1)n}=(\eta^{r}+(1-\eta^{r})N^{m}_{t})       \frac{1}{1+e^{-\sigma^{e} 
(M^{e}_{tn}- \phi^{e})}}
(1-M^{p}_{tn})(1-M^{c}_{tn}) \nonumber\\
((1-\alpha_{tn})-M^{r}_{tn}) \varkappa^{T}+M^{r}_{tn}
\label{eq:5}
\end{eqnarray}
\normalsize
It is affected by the load shedding, mass media, the cooperation, the physical health, and the emotion of the consumers and prosumers. If the emotion ($M^{e}_{tn}$) is lower than the fear or the threshold ($\phi^{e}$), it has no impact on the risk perception \cite{natalie_13_29}. According to the narrowing hypothesis of Fredrickson's broaden-and-build theory \cite{paperasli}, the factor, $[(1-\alpha_{tn})-M^{r}_{ti}]$, measures the tendency of the risk perception to be more or less positive. The relation between the risk perception and the cooperation is provided in \cite{ring1992structuring,fischer2012risk}. The connection between risk perception and physical health is provided in \cite{kim2019effects,stephan2011relation}. 

The optimization given by (1) is subject to a fourth equality constraint, which is the dynamic change of the level of cooperation of consumers and prosumers in CPSS-PE given by
\vspace{-0.3cm}

\footnotesize
\begin{eqnarray}
M^{c}_{(t+1)n}=(\eta^{c}+(1-\eta^{c})N^{m}_{t})  (\frac{1}{1+e^{-\sigma^{c} (M^{e}_{tn}- \phi^{e})}}) \nonumber \\
M^{p}_{tn}M^{a}_{n}[(1- \alpha_{tn}(1-M^{e}_{tn}))   -M^{c}_{tn}] \varkappa^{t} +M^{c}_{tn}.
\label{eq:14}
\end{eqnarray}
\normalsize
It is affected by the emotion, load shedding, and the physical health of consumers and prosumers.  Here, the factor $[(1- \alpha_{tn}(1-M^{e}_{tn}))   -M^{c}_{tn}]$ is based on the narrowing hypothesis of Fredrickson's broaden-and-build theory. The relationship between the fear and the cooperation is provided in \cite{rand2014positive,lebow2005reason,levine2018signaling,kjell2013exploring}. The relation between cooperation and physical health is discussed in \cite{hamalainen2016cross,jensen1994promoting}. According to \cite{fowler2010cooperative,scanlon2011research,ali2013media}, social media influence the level of cooperation among the individuals of a group. 

The optimization given by (1) is subject to a fifth equality constraint, which is the dynamic change of the physical health of consumers and prosumers in CPSS-PE given by

\footnotesize
\begin{equation}
M^{p}_{(t+1)n}=\eta^{p} (\frac{1}{1+e^{-\sigma^{c} 
(M^{e}_{tn}- \phi^{e})}}) 
((1-M^{e}_{tn})\alpha_{tn}-P_{tn}) \varkappa^{t}+M^{p}_{tn}
\label{eq:14b} 
\end{equation}
\normalsize
It is affected by the fear and load shedding of consumers and prosumers. The set of $L_{MP}$=\{$\eta^{r}, \eta^{c}, \eta^{p}$\} includes the mental and physical coefficients. All of the above-mentioned features are assumed to take  values in the interval [0 1].

The optimization given by (1) is subject to a sixth set of equality constraints, which are the power flow equations  using a DC model. They are given by 

\footnotesize
\begin{equation}
P_{nmt}=\frac {\theta_{nt}-\theta_{mt}}{X_{nm}}
\label{eq:14b} 
\end{equation}
\normalsize
Using these power flow equations, we model a set of DERs connected to a bus of the power system that are willing to share their electricity with customers, retailers, private and public organizations connected to other buses of that system.  Their behavior may be viewed as one single group behavior by using a bottom-up approach \cite{Barsade1998} and the equality constraints given by (5)-(11). 

The optimization given by (1) is subject to a seventh equality constraint, which is the power balance between generation and load in the power system expressed as
\vspace{-0.3cm}

\footnotesize
\begin{equation}
\sum_{m} P_{nmt}+P^{mg}_{nt}+P^{der}_{nt}+P^{u}_{nt}= \alpha_{nt} P^{d}_{nt}+\beta_{nt} P^{cl}_{nt}
\label{eq:14b} 
\end{equation}
\normalsize
Note that $(1-\alpha_{nt})$ denotes the fraction of consumers and prosumers that are shed while $(1-\beta_{nt})$ denotes the fraction of the critical loads that are shed. While the effect of the load on the social well-being changes with the seasons or the weather, this effect has not been considered here.

The optimization given by (1) is subject to an eighth set of inequality constraints, which represent the power flow limitations of the transmission lines given by  

\footnotesize
\begin{equation}
- \overline P^{l}_{nm}
\leq
P_{nmt}
\leq
\overline P^{l}_{nm}
\label{eq:14b} 
\end{equation}
\normalsize

The optimization given by (1) is subject to a ninth set of inequality constraints, which represent the limitations of the DERs to generate electricity. They are given  by

\footnotesize
\begin{equation}
0
\leq
P^{der}_{tn}
\leq
 M^{c}_{tn} \overline P^{der}_{n}
\label{eq:14b} 
\end{equation}
\normalsize
The maximum level of sharing of electricity depends on the level of cooperation of the prosumers. The latter may be willing to share their electricity with the customers who do not have electricity during and after a disaster strikes. 

The optimization given by (1) is subject to a tenth set of inequality constraints, which represent the capacities of the microgrids to generate electricity. They are given by

\footnotesize
\begin{equation}
0
\leq
P^{mg}_{nt}
\leq
\overline P^{mg}_{n}
\label{eq:14b} 
\end{equation}
\normalsize
Here, the microgrids and the DERs connected to a bus are assumed to share their electricity with the critical loads such as hospitals, firefighter, police stations, to name a few.  Regarding the sharing of electricity with other customers, we may model more complex behaviors of subsets of DERs and microgrids attached to a bus. As for the data centers, they are assumed to have enough backup generation due to the critical role that they play for smart businesses and government organizations in the modern computing age.  

The optimization given by (1) is subject to a eleventh set of inequality constraints, which are the power plant capacities to generate electricity. They are given by

\footnotesize
\begin{equation}
0
\leq
P^{u}_{nt}
\leq
\overline P^{u}_{n}
\label{eq:14b} 
\end{equation}
\normalsize

The optimization given by (1) is subject to a twelfth set of inequality constraints, which are the voltage angle bounds given by
\vspace{-0.3cm}

\footnotesize
\begin{equation}
-\pi \leq \theta_{nt} \leq \pi
\label{eq:14b} 
\end{equation}
\normalsize

In the proposed model, the severity level of influence of all of the cyber-physical-social factors on each other can be easily modified by adjusting the values given to the  mental and physical coefficients in $L_{MP}$, the weighting factors in $L_{W}$, and the well-being coefficients in $L_{WC}$. \\

\vspace{-0.7cm}

\subsection{Cyber-Physical-Social Dependence in the Multi-Agent-Based Model}
\begin{figure}
\centering
\includegraphics[width=0.6 \columnwidth]{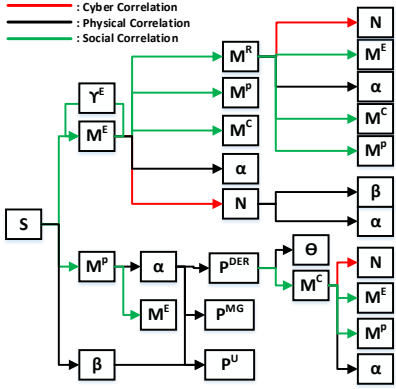}
\caption{Cyber-Physical-Social dependence. Social well-being or community resilience encompasses mental well-being and physical well-being. Load shedding related to both consumers and critical loads influence community resilience. In addition to cyber-physical-social factors, the level of emotion of other connected consumers and prosumers influences that of a particular consumer or prosumer.}
\label{fig:Fig_111}
\vspace{-0.7cm}
\end{figure}

Figure~\ref{fig:Fig_111} displays the CPSS-PE dependence among the characteristics considered in the multi-agent-based model. The social well-being is influenced by the load shedding factor of critical loads, mental well-being, physical well-being of consumers, and prosumers. In this work, we consider the inverse level of fear of consumers and prosumers as their mental well-being. The less fear, the more mental well-being consumers and prosumers have. To model group emotion, we inspire Barsade theory and Fredrickson theory \cite{Barsade1998,Fredrickson2002}. We also consider emotion contagion. The level of empathy among consumers and prosumers influences emotion contagion among them. In addition to the fear propagation, the emotion is affected by the news and information exchanged by mass media platforms,  the level of risk perception, cooperation, physical health, load shedding of consumers and prosumers. The news exchanged in mass media platforms  is directly associated with the load shedding related to consumers, prosumers, and critical loads. Here, we disregard the fake news propagated in mass media platforms. The level of risk perception is affected by the level of emotion, physical health, cooperation, load shedding factor of consumers and prosumers. It is also affected by news and information exchanged by mass media platforms. Load shedding and the level of mental well-being influence the level of physical well-being. The availability of electricity by prosumers, microgrids, and utility affect the load shedding of consumer, prosumers, and critical loads. The availability of power by prosumers is affected by their level of cooperation. The news exchanged thorough mass media platforms , level of fear, physical well-being, and load shedding influence how the prosumers are willing to share electricity.

\vspace{-0.4cm}

\begin{figure}[h]
\centering
\includegraphics[width= 0.7 \columnwidth]{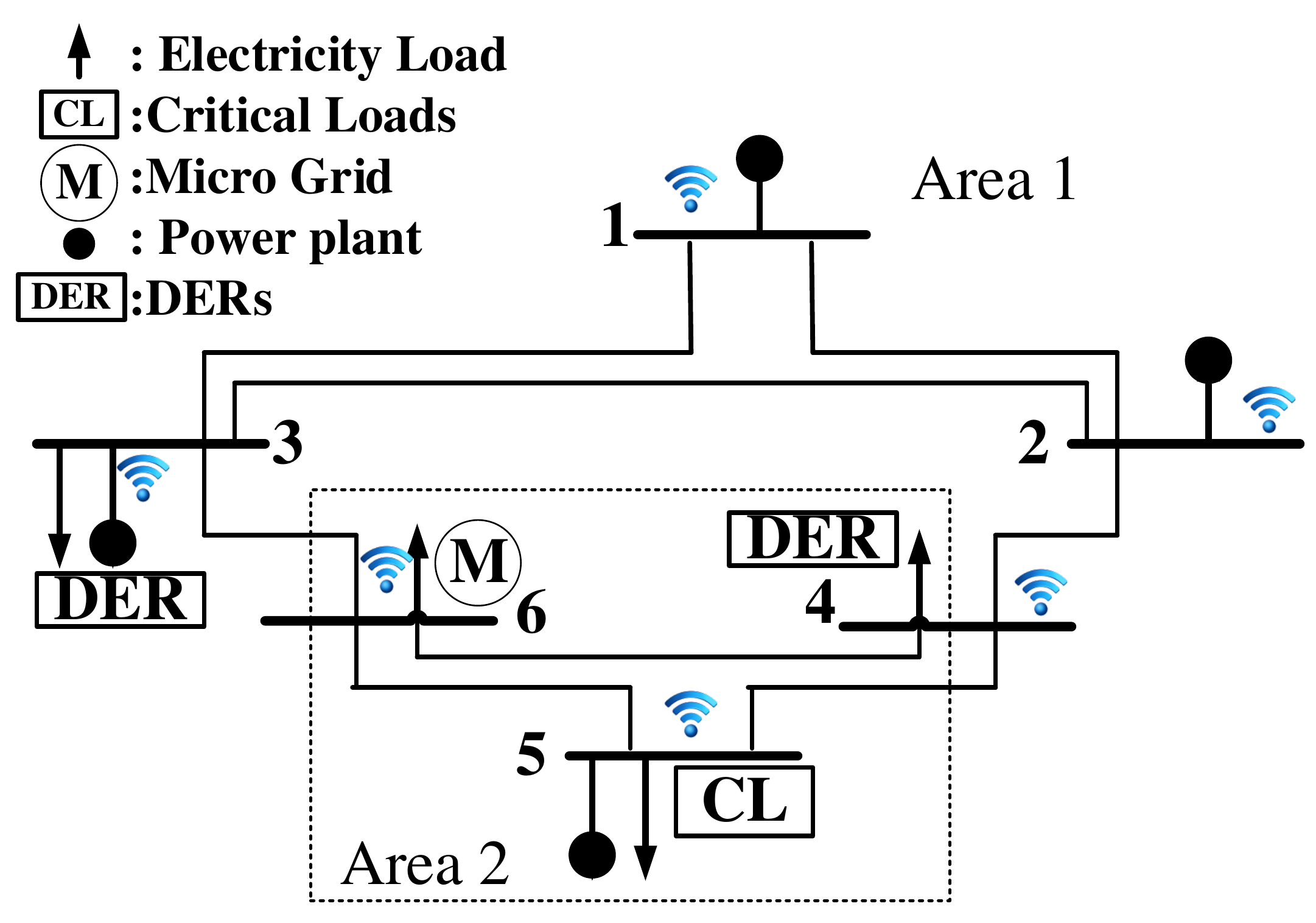}
\vspace{0.5cm}
\setlength{\abovecaptionskip}{-15pt}
\caption{Two-area 6-buses system. There is no congestion between intra-area transmission lines. The inter-area transmission lines have a limited capacity as much as 450 MW (lines 3-6 and 2-4).}
\vspace{-.8cm}
\label{fig:Fig_34a1}
\end{figure}

{\section{Case Study: Two-Area 6-Bus System}}
The first case study is a two-area 6-buses system, as shown in Figure~\ref{fig:Fig_34a1}. This case study aims to provide the results related to the sensitive analysis of different cyber, physical, and social factors shaping community resilience. The data associated with this network are provided in Table~\ref{tab:111}. This table includes the data related to the capacity of power plants, microgrid, and DERs in MW. It also provides the MW demand of consumers and critical loads. The susceptance of transmission lines is assumed to 10 P.U. (100 MW base). It is assumed that all buses have access to the internet and mass media platforms. 
\vspace{-0.4cm}

\begin{table}[H]
\centering
\caption{The data associated with a Two-area 6-buses system (MW).}
\vspace{-0.3cm}
\begin{tabular}{|l| c|c|c|c|c| }
\hline
Bus &	Power Plant & Microgrid & DER & Demand & Critical load  \\
\hline
1 &	650 &  - &-   & -& - \\
\hline	
2 &	297 &  - & -  &  - & - \\
\hline	
3 &	231 & -  &  30 & 750  & - \\
\hline	
4 &	- & -  &  30 & 675  & - \\
\hline	
5 &	100 & -  & -  & 537.5  & 100 \\
\hline	
6 &	- & 50  &  - &  600 & - \\
\hline	
\vspace{-0.4cm}
\label{tab:111}
\end{tabular}
\end{table}

Table~\ref{tab:115} provide the hourly load coefficient for daily consumption. The demand in each hour is obtained by multiplying these coefficients by demand of each bus provided in Table~\ref{tab:111}. It is assumed that all consumers and prosumers follow the same hourly load coefficient trend.
\vspace{-0.3cm}

\begin{table}[H]
\centering
\caption{Hourly Load coefficient for a day. H means hour while LC stands for load coefficient. The data is for 24 hours. }
\vspace{-0.3cm}
\resizebox{0.89\textwidth}{!}{\begin{minipage}{\textwidth}
\begin{tabular}{|l| c|c|c|c|c|c|c|c|c|c|c| }
\hline
H & LC& H&  LC & H&  LC & H&  LC&  H&  LC & H&  LC\\
\hline
1 & 0.23& 2&  0.32 & 3&  0.45 & 4&  0.40&  5&  0.31 & 6&  0.42\\
\hline	
7 & 0.55& 8&  0.21 & 9&  0.40 &10&  0.49& 11&  0.54 &12&  0.55\\
\hline	
13& 0.06&14&  0.18 &15&  0.26 &16&  0.30& 17&  0.37 &18&  0.45\\
\hline	
19& 0.51&20&  0.57 &21&  0.61 &22&  0.84& 23&  1.00 &24&  0.89\\
\hline	
\vspace{-0.4cm}
\label{tab:115}
\end{tabular}
\end{minipage}}
\end{table}

\subsection{Soft Validation of the Proposed CPSS-PE model}
We make a soft validation by verifying the result of the socio-technical power flow model with Case Study 1 provided by \cite{natalie_13}. In the soft validation, only information-seeking behavior, the emotion of fear, and bias are considered in the model. After soft validation, we extend our model to the socio-technical power flow dscribed in the CPSS-PE. To do so, we consider the  cooperation, the empathy, the mass media, the physical well-being of the agents along with the power flow constraints.  
\vspace{-0.4cm}
\subsection{Sensitivity Analysis of Various CPSS-PE Factors in 24 Scenarios}

Table~\ref{tab:aaa} displays the sensitivities of different social, cyber, and physical factors influencing the community resilience. The social factors consist of the level of emotion (fear), cooperation, risk perception, empathy, and physical health. The cyber factor includes the mass media effect factor ($\zeta^{m}$). The physical factors consist of the capacities of the microgrid and of the DERs. Note that in the columns of Table~\ref{tab:aaa} are $M^{E}_{tn}$, $M^{C}_{tn}$, $M^{R}_{tn}$, $M^{A}_{n}$, and $M^{P}_{tn}$, which provide the initial values used for the emotion (fear), cooperation, risk perception, empathy, and physical health, respectively. Here, it is assumed that all the buses have similar initial values. All the outputs of the CPSS-PE in the power system are at an average levels. In total, the results for 24 different scenarios are provided.\par

\textit{Scenarios 1-3 (Changes in the Initial Value of Emotion)}: In these scenarios, the initial value of emotion (fear) is increased from 0.1 to 0.5 to 0.9 while the initial values of the other factors are fixed. This increase results in an increase in the average level of fear. Consequently, the level of risk perception and cooperation is increased while the average level of the physical well-being and community resilience is decreased. An increase in the cooperation reduces the average level of the load shedding. Therefore, less negative news are reported in the mass media platforms. 

\textit{Scenarios 4-6 (Changes in the Initial Value of Cooperation)}: In these scenarios, the initial value of cooperation is increased from 0.1 to 0.5 to 0.9 while the values of the other factors are fixed. This increase results in an increase in the average level of cooperation. Consequently, the amount of load shedding of the consumers, the prosumers, and the critical loads is decreased. Hence, there is less negative news reported in the mass media platforms. In addition, the average level of fear and the risk perception of the consumers and the prosumers are also decreased. Finally, both the physical well-being and the community resilience are increased.

\textit{Scenarios 7-9 (Change in the Initial Value of Risk Perception)}: In these scenarios, the initial value of the risk perception is increased from 0.2 to 0.5 to 0.9 while the values of the other factors are fixed.  This increase results in an increase in the average level of risk perception, fear, and cooperation while the average level of the physical well-being is decreased due to a greater level of fear, stress, and anxiety. Because of an increase in the level of cooperation, the amount of load shedding decreases, resulting in a smaller amount of reported negative news by the mass media platforms. However, the social well-being  and the community resilience are reduced. 

\textit{Scenarios 10-12 (Change in the Level of Empathy)}: In these scenarios, the initial value of empathy is increased from 0.1 to 0.5 to 0.9 while the values of the other factors are fixed. This increase  results in an increase in the average level of empathy. Consequently, the amount of load shedding and related negative news in the mass media platforms is decreased. Also, the average level of fear along with the risk perception of the consumers and the prosumers  decline. Finally, both the physical well-being and the community resilience increase.

\textit{Scenarios 13-15 (Change in the Initial value of the Physical Health)}: In these scenarios, the initial value of the physical health of people is increased from 0.1 to 0.5 to 0.9 while the values of the other factors are fixed. This increase results in an increase of the average level of the physical well-being, mental well-being, and cooperation. The amount of load shedding and negative news reported by the mass media platforms declines. Therefore, the community resilience improves.

\textit{Scenarios 16-18 (Change in the Mass Media Effect Factor)}: In these scenarios, the level of the social media effect factor is increased from 0.1 to 0.5 to 1 while the values of the other factors are  fixed. This increase results in an increase in the negative news and  the average level of fear. Hence, the average level of cooperation and risk perception increases. On the other hand, the amount of load shedding and physical well-being decreases. In addition, because of the high effect of the mass media on the propagation of negative news, the average level of community resilience decline.

\textit{Scenarios 19-21 (Change in the Total DER Capacity)}: In these scenarios, the total DER capacity is increased from 0 to 60 to 200 MW while the values of the other factors are fixed.  This increase results in  a decrease of the load shedding, especially of the  critical loads. The negative news reported by the mass media is decreased. In addition, the average level of fear, cooperation, risk perception is decreased while that of the physical health is increased. As a result, the community resilience is enhanced. 

\textit{Scenarios 21-24 (Change in the Total Microgrid Capacity)}: In these scenarios, the total microgrid capacity is increased from 0 to 50 to 300 MW while the values of the other factors are fixed.  This increase results in a decrease in the load shedding, resulting in an increase of the community resilience. The levels of the social well-being factors and negative news have the same trends as those of Scenarios 19-21. \\
\vspace{-0.8cm}

\tiny
\begin{table*}[h!]
\label{tab:aaa}
\caption{The results of our community resilience optimization method subject to power flow constraints in CPSS-PE. All the results are at an average level for 24 hours. CR, $L^{\alpha}$ \& $L^\beta$ stand for community resilience, load shedding of consumers and prosumers, and load shedding of critical loads, respectively. DER and microgrid capacities are in MW. }
\resizebox{0.9\textwidth}{!}{\begin{minipage}{\textwidth}
\begin{tabular}{|c|l|l|l|l|l|l|l|l|l|l|l|l|l|l|l|l|l|}
\hline
\hline
\multicolumn{2}{ |c| }{CPSS in power engineering} & \multicolumn{8}{ |c| }{Inputs of community resilience optimization in CPSS-PE} & \multicolumn{8}{ |c| }{Outputs of community resilience optimization in CPSS-PE} \\
\hline \hline
System &Change Factor& $M^{E}_{tn}$ & $M^{C}_{tn}$ & $M^{R}_{tn}$ & $M^{A}_{n}$ & $M^{P}_{tn}$ & $\zeta^{m}$& $\overline P^{DER}$ & $\overline P^{MG}$ & CR(S) & $L^{\alpha}$ & $L^\beta$ & $M^{E}$ & $M^{C}$ & $M^{R}$ & $M^{P}$ & $N^{M}$ \\ \hline
\hline
\multirow{15}{*}{Social} &Emotion & 0.1 & 0.5 & 0.5 & 1 & 0.5 & 1 & 60 & 50 & 0.655 & 0.632 & 0.153 & 0.616 & 0.669 & 0.593 & 0.384 & 0.651 \\ 
\cline{2-18}
&Emotion & 0.5 & 0.5 & 0.5 & 1 & 0.5 & 1 & 60 & 50 & 0.634 & 0.632 & 0.149 & 0.686 & 0.713 & 0.62 & 0.347 & 0.649 \\ 
\cline{2-18}
&Emotion & 0.9 & 0.5 & 0.5 & 1 & 0.5 & 1 & 60 & 50 & 0.614 & 0.631 & 0.148 & 0.75 & 0.727 & 0.625 & 0.334 & 0.649 \\ 
\cline{2-18}
&Cooperation & 0.5 & 0.1 & 0.5 & 1 & 0.5 & 1 & 60 & 50 & 0.604 & 0.635 & 0.171 & 0.751 & 0.502 & 0.691 & 0.341 & 0.657 \\ 
\cline{2-18}
&Cooperation & 0.5 & 0.5 & 0.5 & 1 & 0.5 & 1 & 60 & 50 & 0.634 & 0.632 & 0.149 & 0.686 & 0.713 & 0.62 & 0.347 & 0.649 \\ 
\cline{2-18}
&Cooperation & 0.5 & 0.9 & 0.5 & 1 & 0.5 & 1 & 60 & 50 & 0.665 & 0.627 & 0.131 & 0.606 & 0.933 & 0.526 & 0.363 & 0.642 \\ 
\cline{2-18}
&Risk Perception& 0.5 & 0.5 & 0.2 & 1 & 0.2 & 1 & 60 & 50 & 0.656 & 0.632 & 0.149 & 0.609 & 0.692 & 0.38 & 0.365 & 0.65 \\ 
\cline{2-18}
&Risk Perception& 0.5 & 0.5 & 0.5 & 1 & 0.5 & 1 & 60 & 50 & 0.634 & 0.632 & 0.149 & 0.686 & 0.713 & 0.62 & 0.347 & 0.649 \\ 
\cline{2-18}
&Risk Perception& 0.5 & 0.5 & 0.9 & 1 & 0.9 & 1 & 60 & 50 & 0.609 & 0.631 & 0.148 & 0.767 & 0.719 & 0.912 & 0.341 & 0.649 \\ 
\cline{2-18}
&Empathy & 0.5 & 0.5 & 0.5 & 0.1 & 0.5 & 1 & 60 & 50 & 0.616 & 0.634 & 0.17 & 0.726 & 0.53 & 0.661 & 0.345 & 0.655 \\ 
\cline{2-18}
&Empathy & 0.5 & 0.5 & 0.5 & 0.5 & 0.5 & 1 & 60 & 50 & 0.625 & 0.633 & 0.158 & 0.705 & 0.628 & 0.641 & 0.346 & 0.652 \\ 
\cline{2-18}
&Empathy & 0.5 & 0.5 & 0.5 & 1 & 0.5 & 1 & 60 & 50 & 0.634 & 0.632 & 0.149 & 0.686 & 0.713 & 0.62 & 0.347 & 0.649 \\ 
\cline{2-18}
&Physical Health & 0.5 & 0.5 & 0.5 & 1 & 0.1 & 1 & 60 & 50 & 0.524 & 0.635 & 0.167 & 0.785 & 0.561 & 0.711 & 0.074 & 0.655 \\ 
\cline{2-18}
&Physical Health & 0.5 & 0.5 & 0.5 & 1 & 0.5 & 1 & 60 & 50 & 0.634 & 0.632 & 0.149 & 0.686 & 0.713 & 0.62 & 0.347 & 0.649 \\ 
\cline{2-18}
&Physical Health & 0.5 & 0.5 & 0.5 & 1 & 0.9 & 1 & 60 & 50 & 0.744 & 0.63 & 0.14 & 0.587 & 0.772 & 0.537 & 0.647 & 0.647 \\ \hline \hline
\multirow{3}{*}{Cyber} & Mass Media & 0.5 & 0.5 & 0.5 & 1 & 0.5 & 0.1 & 60 & 50 & 0.666 & 0.633 & 0.16 & 0.517 & 0.597 & 0.566 & 0.396 & 0.065 \\ 
\cline{2-18}
&Mass Media & 0.5 & 0.5 & 0.5 & 1 & 0.5 & 0.5 & 60 & 50 & 0.649 & 0.632 & 0.153 & 0.597 & 0.663 & 0.601 & 0.359 & 0.325 \\ 
\cline{2-18}
&Mass Media & 0.5 & 0.5 & 0.5 & 1 & 0.5 & 1 & 60 & 50 & 0.634 & 0.632 & 0.149 & 0.686 & 0.713 & 0.62 & 0.347 & 0.649 \\ \hline \hline
\multirow{6}{*}{Physical} &DER Capacity& 0.5 & 0.5 & 0.5 & 1 & 0.5 & 1 & 0 & 50 & 0.59 & 0.642 & 0.231 & 0.697 & 0.719 & 0.624 & 0.343 & 0.674 \\ 
\cline{2-18}
&DER Capacity& 0.5 & 0.5 & 0.5 & 1 & 0.5 & 1 & 60 & 50 & 0.634 & 0.632 & 0.149 & 0.686 & 0.713 & 0.62 & 0.347 & 0.649 \\ 
\cline{2-18}
&DER Capacity& 0.5 & 0.5 & 0.5 & 1 & 0.5 & 1 & 200 & 50 & 0.694 & 0.601 & 0.042 & 0.66 & 0.695 & 0.606 & 0.36 & 0.601 \\ 
\cline{2-18}
&MG Capacity& 0.5 & 0.5 & 0.5 & 1 & 0.5 & 1 & 60 & 0 & 0.571 & 0.646 & 0.266 & 0.701 & 0.722 & 0.627 & 0.341 & 0.683 \\ 
\cline{2-18}
&MG Capacity& 0.5 & 0.5 & 0.5 & 1 & 0.5 & 1 & 60 & 50 & 0.634 & 0.632 & 0.149 & 0.686 & 0.713 & 0.62 & 0.347 & 0.649 \\ 
\cline{2-18}
&MG Capacity& 0.5 & 0.5 & 0.5 & 1 & 0.5 & 1 & 60 & 300 & 0.717 & 0.574 & 0.006 & 0.638 & 0.685 & 0.592 & 0.366 & 0.568 \\ \hline
\hline
\end{tabular}
\vspace{-0.6cm}
\end{minipage}}
\end{table*}

\normalsize

\section{Case Study 2: the Modified IEEE RTS 24-Bus System}
\begin{figure}[h]
\centering
\includegraphics[width= 0.6 \columnwidth]{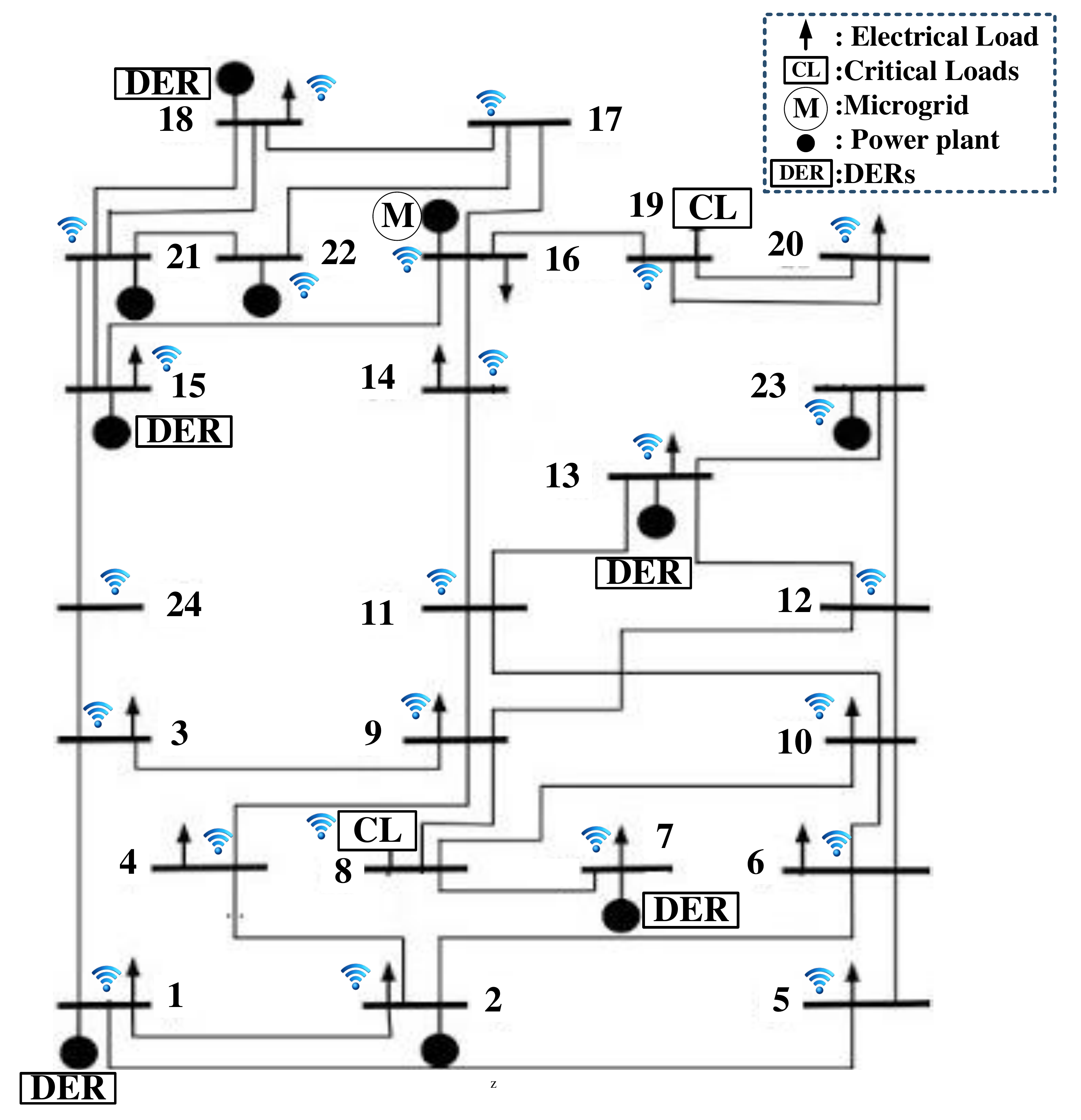}
\vspace{0.4cm}
\setlength{\abovecaptionskip}{-15pt}
\caption{The one-line diagram of the modified IEEE RTS 24-bus system. It is assumed that all the buses have access to the internet and to the mass media platforms. }
\vspace{-0.5cm}
\label{fig:Fig_34a}
\end{figure}

A modified IEEE RTS 24-bus system is used to implement the proposed socio-technical power flow in the CPSS-PE, as it is displayed in Figure~\ref{fig:Fig_34a}. Bus 16 has a microgrid with a capacity of 310 MW.  Additionally, the total capacities of the DERs connected to Buses 1, 7, 13, 15, and 18 are 50, 50, 100, 50, and 100 MW, respectively. It is assumed that there are two critical loads connected to Buses 8, and 19 of 426 MW and 451 MW, respectively. An initial level of 0.5 is assumed for the cooperation, emotion (fear), risk perception, and physical health of all buses, including consumers, prosumers, microgrid owners, critical loads, and utilities. In addition, to prevent making the problem complex, we assume that there is an empathy level of 1 between two buses if there is a transmission line between them. The socio-technical power flow algorithm is executed for 24 hours. It is assumed that the generation units located at Buses 21 and 23 are turned off since hour 5. Moreover, the generation units located at Buses 1, 2, 7, 13, 15, and 16 are turned off since hour 14. All the DERs and microgrids are connected to the power system for the whole time.  
\vspace{-0.7cm}

\begin{figure}[h]
\centering
\includegraphics[width=  \columnwidth]{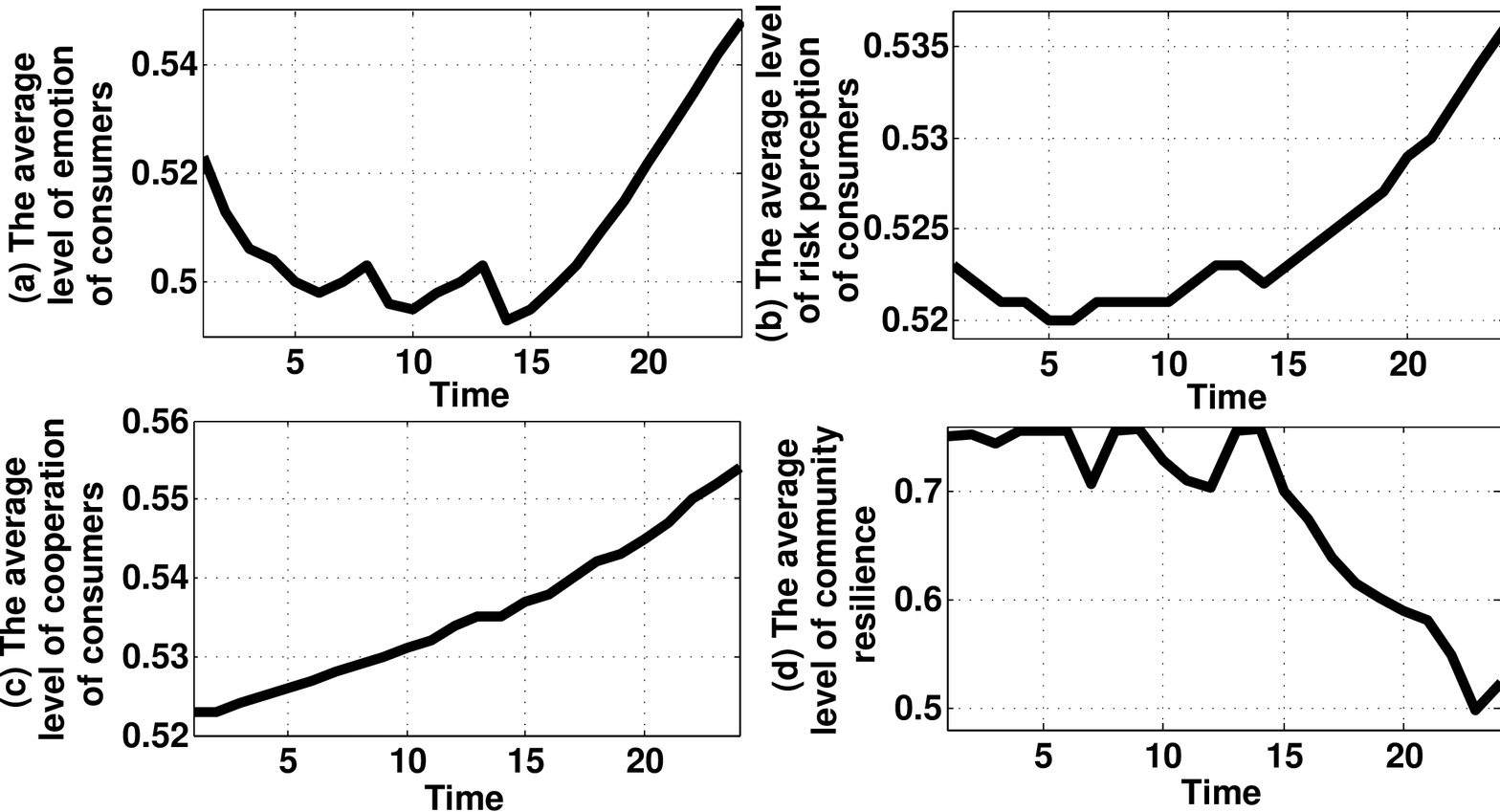}
\setlength{\abovecaptionskip}{-15pt}
\caption{Dynamic change of social behavior of the consumers, prosumers, and the whole community of the modified IEEE RTS 24-bus system; (a) The average level of emotion per hour; (b) The average level of risk perception per hour; (c) The average level  of cooperation per hour; (d) The average level  of community resilience per hour.
}
\vspace{-0.7cm}
\label{fig:Fig_34bb}
\end{figure}

\begin{figure}[h]
\centering
\includegraphics[width= \columnwidth]{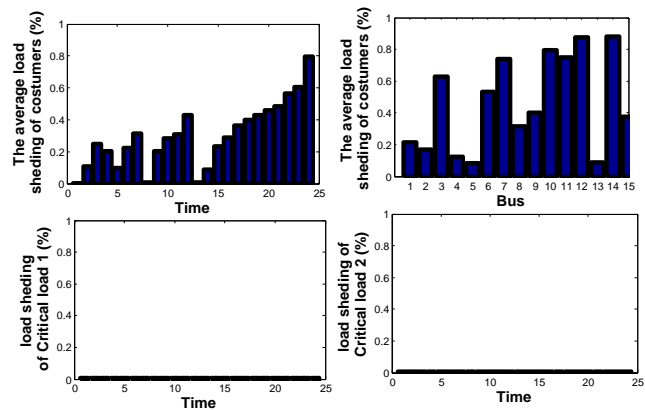}
\setlength{\abovecaptionskip}{-15pt}
\caption{Load shedding experienced by the consumers, the prosumers, and the critical loads obtained by running a socio-technical power flow algorithm on the modified IEEE RTS 24-bus system; (a) the average load shedding experienced by consumers and prosumers per hour; (b) the average load shedding experienced by consumers and prosumers per bus; (c) the load shedding experienced by the critical load at Bus 8 per hour; (d) the load shedding experienced by the critical load at Bus 19 per hour.}
\vspace{-0.7cm}
\label{fig:Fig_34cc}
\end{figure}

Figures~\ref{fig:Fig_34bb} and ~\ref{fig:Fig_34cc} provide the result of the socio-technical power flow in CPSS-PE.  Figure~\ref{fig:Fig_34bb} displays the dynamic change in the level of emotion, risk perception, cooperation of costumers, and prosumers, in addition to the dynamic change in the level of community resilience of the entire society connected to the IEEE RTS 24-bus system. The level of emotion (fear) of consumers and prosumers depends on the emotion contagion, cooperation, load shedding, and physical health, to name a few. The level of emotion fluctuates from hour 1 to hour 14. Afterward, the levels of fear of the  consumers and the prosumers increase significantly due to the high level of load shedding.  Furthermore, because some generating units are turned off since hour 14, the consumers and prosumers experience a high level of risk of not being supplied with electricity. This situation prompts them to cooperate by sharing electricity in case of a shortage. Because the community resilience is highly intertwined with the critical loads in the CPSS-PE, it decreases noticeably since hour 14 due to power generation shortage. The average level of community resilience of the entire society connected to the  IEEE RTS 24-bus system attains 0.682. The highest level of community resilience occurs at hour 13 since the load shedding is at its lowest level.

Figure~\ref{fig:Fig_34cc} presents the results of the load shedding experienced by the consumers at Buses 2 to 6 and at Buses 9, 10, 14, 20 and the prosumers at Buses 1, 7, 13, 15, and 18, and the critical loads at Buses 8 and 19 in CPSS-PE. Understandably, there is no load shedding in the buses without a demand. The average levels of load shedding experienced by the critical loads at Bus 8 and 19 amount to 0.275 and 0.013, respectively, yielding a total average of 0.144. The average levels of load shedding experienced by the consumers and the prosumers amount to 0.401.
\vspace{-0.7cm}

\section{Conclusions}

In this paper, we developed a community resilience optimization method subject to power flow constraints in CPSS-PE. The socio-technical power flow model includes the social constraints, i.e., the dynamic change of the level of emotion, risk perception, cooperation, and physical well-being of consumers and prosumers. We also examine the effect of critical loads on the social well-being. In addition to the  social constraints, we include in the model the cyber constraints and the physical constraints. The proposed model is implemented in two  different case studies, i.e., a two-area 6-bus system and a modified IEEE RTS 24-bus system. The result of a sensitive analysis carried out on the cyber-physical-social factors that characterize the community resilience can be summarized as follows:
\begin{itemize}
\item In the social aspect, an increase in the initial value of the emotion, risk perception of the society under study because of the culture and the previous experience, to name a few, results in the decrease of the level of both the load shedding and the community resilience. On the other hand, an increase in the initial value of cooperation, empathy, and physical health results in the decrease of the level of the load shedding and an increase in the level of the community resilience.  
\item In the cyber aspect, an increase in the social media platform effect factor leads to a decrease in the level of both the load shedding and the community resilience. 
\item In the physical aspect, the larger the installed capacity of the  microgrids and DERs, the smaller the level of load shedding and the larger the level of community resilience.
\end{itemize}
We also provide the dynamic effect of the load shedding experienced by the consumers, prosumers, and the critical loads on the social behavior. The results show that the prosumers cooperate to share electricity since they face a power shortage.  As a future work,  the investment in microgrids to enhance the community resilience will be investigated. Sharing electricity is useful for both economic and resiliency aspects. this may be achieved by installing one microgrid per cluster of critical loads, such as hospitals, instead of providing each of them with a backup generator.  
\vspace{-0.5cm}

\addcontentsline{toc}{section}{References}
\bibliographystyle{IEEEtran}
\bibliography{neighbourhood}

\end{document}